\documentclass[
twocolumn,
amsfonts,
showpacs,
superscriptaddress,
amsmath,
amssymb,
aps,
longbibliography,
prl,
noeprint
]{revtex4-2}
\usepackage[normalem]{ulem}

\usepackage{graphicx}
\usepackage{bm}
\usepackage[dvipsnames]{xcolor}
\usepackage[english]{babel}
\usepackage{dsfont}
\usepackage{comment}
\usepackage[caption=false]{subfig}
\usepackage[colorlinks,
            linkcolor=BrickRed,%
            citecolor=MidnightBlue,%
            urlcolor=MidnightBlue]{hyperref}

\newcommand{\ch}[1]{{\color{black}  #1}}

\newcommand{\be}{\begin{equation}}
\newcommand{\ee}{\end{equation}}
\newcommand{\uu}{\underline{u}}

\newcommand{\Tx}{{x}}
\newcommand{\Tt}{{t}}

\newcommand{\veff}{v^\eff}
\newcommand{\bveff}{\bar{v}^\eff}

\newcommand{\dr}{{\mathrm{dr}}}

\newcommand{\para}[1]{ \noindent\emph{\textbf{#1---}}}

\newcounter{ls}

\newcommand{\bc}{\begin{center}}
\newcommand{\ec}{\end{center}}
\def\ba#1{\begin{array}{#1}\displaystyle}
\newcommand{\ea}{\end{array}}

\newcommand{\beq}{\begin{equation}}
\newcommand{\eeq}{\end{equation}}
\newcommand{\beqa}{\begin{eqnarray}}
\newcommand{\eeqa}{\end{eqnarray}}

\newcommand{\n}{\nonumber\\}
\newcommand{\bi}{\begin{itemize}}

\newcommand{\p}{\partial}

\newcommand{\ii}{{\rm i}}

\newcommand{\dd}{{\rm d}}

\newcommand{\eff}{{\rm eff}}

\begin{document}

\title{Anomalous hydrodynamic fluctuations in the quantum XXZ spin chain}
\author{Takato Yoshimura}
\email{takato.yoshimura@physics.ox.ac.uk}
\affiliation{Department of Mathematics, King’s College London, Strand, London WC2R 2LS, U.K.}
\affiliation{All Souls College, Oxford OX1 4AL, U.K.}
\affiliation{Rudolf Peierls Centre for Theoretical Physics, University of Oxford, 1 Keble Road, Oxford OX1 3NP, U.K.}

\author{\v{Z}iga Krajnik}
\email{ziga.krajnik@nyu.edu}
\affiliation{Department of Physics, New York University, 726 Broadway, New York, NY 10003, USA}
\author{Alvise Bastianello}
\affiliation{CEREMADE, CNRS, Universit\'e Paris-Dauphine, Universit\'e PSL, 75016 Paris, France}
\author{Enej Ilievski}
\affiliation{Department of Physics, Faculty of Mathematics and Physics, University of Ljubljana, Jadranska 21, SI-1000 Ljubljana, Slovenia}

\begin{abstract}
{The quantum XXZ spin-1/2 chain features non-Gaussian spin current fluctuations in the regime of easy-axis anisotropy.
Using ballistic macroscopic fluctuation theory, we derive the exact \ch{asymptotic} probability distribution of typical spin-current fluctuations in thermal equilibrium \ch{at zero average magnetization}. The obtained nested Gaussian distribution is fully characterized by its variance which we analytically relate to the spin diffusion constant and static spin susceptibility, and compare our prediction with numerical simulations.
By unveiling how the same mechanism which leads to anomalous charge current fluctuations in single-file systems manifests itself in the XXZ chain, our approach establishes the universal hydrodynamic origin of the observed anomalous fluctuations.}
\end{abstract}

\maketitle

\para{Introduction.} Quantifying macroscopic fluctuations in interacting many-body systems is one of the major challenges in statistical mechanics. In spite of the inherent complexity of their microscopic dynamics, the emergent collective behavior of observables at macroscopic scales---such as densities or currents of conserved charges---typically exhibits an astounding degree of simplicity complying with central-limit-type behavior. The robustness of Gaussian statistics in equilibrium is one of the most remarkable features of statistical mechanics. Departures from Gaussian universality are rare yet particularly relevant in one dimensional systems where strong correlations, additional conservation laws, or dynamical constraints can qualitatively alter the behavior of fluctuations.

By extending the standard path-integral approach to stochastic hydrodynamics~\cite{MSR}, modern macroscopic fluctuation theory (MFT)~\cite{MFT} has solidified its status as a powerful and versatile tool for describing statistical properties of fluctuating macroscopic observables in generic, diffusive systems in a wide array of physical scenarios.  As such, MFT offers a universal way to characterize the distribution of time-integrated currents, which are usually Gaussian.
\ch{It therefore came as a surprise that certain interacting systems exhibit \emph{anomalous} (i.e. non-Gaussian) fluctuations of time-integrated currents of ``nested Gaussian'' form
\be
     \mathcal{P}_\mathrm{typ}(j)= \frac{1}{2\pi \sigma}\int_\mathbb{R} \frac{\dd x}{\sqrt{|x|}}\,\exp\left[-\frac{x^2}{2\sigma^2}-\frac{j^2}{2|x|}\right].\label{eq:typ_dist}
\ee}
Here\ch{, the} variable $j$ denotes the time-integrated  current  rescaled to the typical diffusive timescale $ J(t)\simeq j\,t^{1/4}$ at large times, while $\sigma$ is a non-universal parameter.

\begin{figure}
    \centering
\includegraphics[width=\columnwidth]{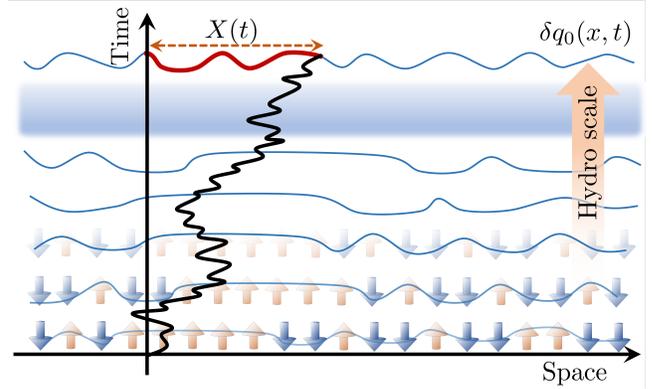}
\caption{\textbf{Origin of anomalous fluctuations of the integrated spin current.} At the hydrodynamic scale, the magnetization of microscopic spin excitations is captured by a large-scale classical fluctuating field $\delta q_0(x,t)$ (blue curves). 
    Anomalous fluctuations stem from the nesting of two sources of Gaussian fluctuations. Transported magnetization equals the fluctuating magnetization in the interval of length $|X(t)|$ (highlighted in red over the fluctuating magnetization), where $X(t)$ (black curve) is the trajectory of a giant magnon that itself fluctuates due to scattering with other modes.}
\label{fig:cartoon}
\end{figure}

\ch{The distribution \eqref{eq:typ_dist} was first numerically observed in Ref.~\cite{Krajnik2022_1} for the magnetization current in integrable classical spin chains, in the so-called easy-axis regime which energetically favors ferromagnetic order. A phenomenological picture was later proposed in Ref.~\cite{SarangKhemani2024} (see also Refs.~\cite{gopalakrishnan2024nongaussian,Kormos2022}) in the quantum XXZ spin chain, valid deep in the easy-axis and in the high-temperature limit. A detailed and quantitative understanding for the emergence of Eq.~\eqref{eq:typ_dist} away from this limit has nevertheless remained elusive, both in the quantum and classical spin chains.
Interestingly, the same nested Gaussian form, Eq.~\eqref{eq:typ_dist}, has also been found in charged single-file systems, first from an ab-initio microscopic derivation~\cite{Krajnik2022,Krajnik_Schmidt_Pasquier_Prosen_Ilievski_2024,krajnik2024singlefile,krajnik2025} and subsequently also using hydrodynamic methods~\cite{Yoshimura2025,Yoshimura2025_2}. We also note that a similar nested Gaussian distribution was found in different contexts, e.g. the position distribution of a particle in a system with nonreciprocal interactions \cite{Basu2024}, albeit through completely different mechanism.

Investigating whether hydrodynamics could quantitatively explain anomalous fluctuations in the quantum XXZ chain is a compelling question, which we answer positively in  this Letter. 
We employ ballistic macroscopic fluctuation theory (BMFT)~\cite{BMFT,Doyon2023longrange,Hubner2025,kethepalli2025,Urilyon2026,mukherjee2026}--- an extension of MFT to systems with ballistic modes, including integrable models described by generalized hydrodynamics (GHD)~\cite{GHD1,GHD2,Bastianello2022,Doyon2025_perspective}. Our approach can be readily adapted to other integrable quantum spin chains~\cite{Ilievski2017} and classical magnets (e.g.~the Landau-Lifshitz magnet~\cite{bastianello2024landaulifschitz}).

Anomalous current fluctuations arise from a simple hydrodynamic mechanism analogous to that in charged single-file systems~\cite{Yoshimura2025}. In essence, the time-integrated spin current density at a fixed position (representing the transferred magnetization across the junction) receives two contributions: one from the normally-distributed fluctuating magnetization within a region and another from the dynamical boundary $X(t)$ of that region. The position $X(t)$ is also normally-distributed, and described by a fluctuating hydrodynamic mode with zero mean velocity driven by other ballistic hydrodynamic modes (see Fig.~\ref{fig:cartoon}), and thus experiencing \emph{convective diffusion} \cite{Medenjak2020}. 
Furthermore, we link $X(t)$ to trajectories of so-called {\it giant magnons} \cite{DeNardis2019}, i.e.~bound states of a large number of fundamental magnons.

By reconciling the hydrodynamic descriptions of the quantum XXZ spin chain and charged single-file models, our approach provides a unified framework for studying anomalous charge fluctuations in related systems, avoiding fully microscopic derivations which are in most cases, including the XXZ spin chain, presently out of reach.
}

\para{Fluctuating hydrodynamics of the XXZ chain.} We consider the quantum spin$-1/2$ XXZ chain in the thermodynamic limit, with the Hamiltonian
\be\label{eq_Ham}
\hat{H}=\frac{1}{4}\sum_{\ell} \hat{\sigma}^x_\ell\hat{\sigma}^x_{\ell+1}+\hat{\sigma}^y_\ell\hat{\sigma}^y_{\ell+1}+\Delta \hat{\sigma}^z_\ell\hat{\sigma}^z_{\ell+1}\,, 
\ee
where $\hat{\sigma}^{x,y,z}$ are Pauli matrices and \ch{we focus on the easy-axis regime with anisotropy $\Delta > 1$.} 
We consider \ch{thermal ensembles} $\hat{\varrho}\propto e^{-\beta \hat{H}+\mu \hat{S}^z}$ \ch{with inverse temperature $\beta$ and chemical potential $\mu$,} where $\hat{S^z}=\frac{1}{2}\sum_\ell\hat{\sigma}^z_\ell$\ch{ , and eventually focus on the half-filled limit $\mu\to 0$}. As is customary in the XXZ chain where the $z-$magnetization is the first of an infinite hierarchy of conserved charges, we define $\hat{q}_{0;\ell}=\frac{1}{2}\hat{\sigma}_\ell^z$, and the spin current $\hat{\jmath}_{0;\ell}$ satisfies the continuity equation $\partial_t\hat{q}_{0;\ell}+\hat{\jmath}_{0;\ell+1}-\hat{\jmath}_{0;\ell}=0$. From the continuity equation, the integrated current $\hat{J}(t)=\int_0^t\dd t'\hat{\jmath}_{0;\ell}(t')$ is expressed as the transferred magnetization across the origin $\hat{J}(t)=\sum_{\ell\ge 0}[\hat{q}_{0;\ell}(t)-\hat{q}_{0;\ell}(0)]$.
In the following, we omit the site label $\ell$ whenever not crucial.

Due to integrability, the model hosts infinitely-many species of long-lived excitations undergoing completely elastic (i.e.~non-diffractive) scattering~\cite{Doyon_GHD_lecture}. These quasiparticles are spin waves (magnons) above a ferromagnetic state and bound states thereof (also called ``strings"), specified by an internal index $n\in\mathbb{N}$. Their dispersion relations are determined by the energies $E_{u,n}$ and momenta $p_{u,n}$, conveniently parametrized in terms of the rapidity variable $u\in[-\pi/2,\pi/2)$. The bare magnetization of a bound state equals $m_n=n$ and counts the number of flipped spins within a string. 
\ch{We conveniently group these labels into a single index $\uu\equiv(u,n)$.}

We proceed by outlining the hydrodynamics in terms of these quasiparticles, while the thermodynamics is reported in End Matter. At hydrodynamic scales we coarse grain the lattice spacing $\ell$ to a continuous spatial coordinate $x$.
At the Euler scale (i.e.~$x,t$ are taken to infinity while their ratio remains finite), one introduces the local densities of excitations (quasiparticles) $\rho_{\uu}(x,t)$, which evolve according to the GHD equation~\cite{GHD1,GHD2}
\be\label{eq_ghd}
\partial_t \rho_{\uu}+\partial_x(\veff_{\uu}\rho_{\uu})=0\, .
\ee
The effective velocities $\veff_{\uu}[\rho]$ describe the ballistic propagation of quasiparticles and depend nonlinearly on all quasiparticle densities, which we collectively denote by $\rho$ without the subscript $\uu$.  
Explicit expressions for the dispersion relations and scattering shifts are reported in End Matter.
Equation~\eqref{eq_ghd} was first proposed to describe the hydrodynamic evolution of the average (i.e.~non-fluctuating) quasiparticle densities evolving on a homogeneous background.

The key idea of BMFT is that Eq.~\eqref{eq_ghd} also captures the time-evolution of {\it hydrodynamic} fluctuations, and in particular {\it large} (i.e.~Euler-scale) fluctuations. \ch{Hence,} Eq.~\eqref{eq_ghd} also holds for {\it fluctuating} quasiparticle densities at the Euler scale, which we still denote by $\rho_{\underline{u}}$, with the fluctuating velocities $\veff_{\uu}[\rho]$~\cite{BMFT,Doyon2023longrange}. The idea was recently extended to describe the time-evolution of {\it typical} fluctuations and was used in evaluating the typical charge current distribution in cellular automata \cite{Yoshimura2025,Yoshimura2025_2}. \ch{Indeed, Ref.~\cite{Medenjak2020} already proposed (and later Refs.~\cite{gopalakrishnan2024nongaussian,Yoshimura2025,McCulloch2025,Yoshimura2025_2} further explored) that diffusive-scale dynamics in integrable systems takes place due to convective propagation of diffusive fluctuations, indicating that convective diffusion is the only mechanism of diffusion and there is no normal diffusion in integrable systems. Our approach therefore naturally incorporates this idea within the framework of BMFT.}

The characterization of initial fluctuations is straightforward: 
since $\rho_{\uu}$ describe coarse-grained quasiparticle densities, the central limit theorem applies and their typical fluctuations are Gaussian with known variances~\cite{Doyon_GHD_lecture}.  Quasiparticle densities in generic integrable systems fully specify an equilibrium macrostate, and charge density fluctuations thus inherit their Gaussian structure. The easy-axis quantum XXZ spin chain is however exceptional in this regard: since the (ferromagnetic) quasiparticle vacuum state is doubly-degenerate, an additional bit $s = \pm1 $, prescribing the vacuum polarization direction, is required to uniquely determine a macrostate. 

As already demonstrated in Ref.~\cite{Piroli2017} (see also Ref.~\cite{Bastianello2022_AB}), properly incorporating the information carried by $s$ is crucial for obtaining the complete hydrodynamic description of magnetization. 
\ch{In equilibrium at $\langle \hat{q}_{0}\rangle=0$, however, $s$ ceases to be a good quantum number due to restoration of the spin-reversal parity symmetry. As we detail out below, fluctuations of magnetization are instead linked to those of giant magnons.}

Before diving into the derivation, it is worthwhile to outline our strategy. The first step is to characterize the typical fluctuations of $\rho_{\uu}$ at time $t=0$. Next, these are connected to those of magnetization and finally, the latter are propagated in time using Eq.~\eqref{eq_ghd} and typical fluctuations of the integrated current computed.\\
\para{Fluctuations of quasiparticle densities}
Since the dynamical exponent of spin transport at half-filling is $z=2$, we introduce a scaling parameter $T \gg 1$ and rescaled coordinates $(\sqrt{T}\Tx,T \Tt)$.
Using quasiparticle density operators $\hat{\rho}_{\underline{u}}$ 
\cite{Ilievski2017-2,BMFT,Yoshimura2025} 
we introduce quasiparticle density fields $\rho_{\underline{u}}(x,t)$,  identified with $\hat{\rho}_{\underline{u}}(\sqrt{T}x,T t)$.
While $\rho_{\underline{u}}(x,t)$ are quantum operators, \ch{like in BMFT, we regard them}
as {\it classical} fluctuating fields: this identification holds in a weak sense when evaluated in correlation functions at the diffusive scale, see Fig.~\ref{fig:cartoon}. To ease the notation, we hereafter still use  $\rho_{\uu}(x,t)$ to denote these classical fluctuating fields.

They still satisfy the {\it Euler} GHD equation
$\p_t\rho_{\underline{u}}+\sqrt{T}\p_x(\veff_{\underline{u}}[\rho]\rho_{\underline{u}})=0$, contributing to typical fluctuations at late times. The appearance of $\sqrt{T}$ in the above equation is due to the mismatch of the scaling we take (diffusive) and that of the underlying dynamics (Euler).
It is more convenient to deal with filling functions defined by $\vartheta_{\uu}=\rho_{\uu}/\rho^\mathrm{tot}_{\uu}$ where total densities $\rho^\mathrm{tot}_{\uu}$ represent the densities of available states (see End Matter). Following the same procedure as above, we also treat $\vartheta_{\uu}(x,t)$ as classical fluctuating fields, which propagate according to $\partial_t\vartheta_{\uu}+\sqrt{T}v^\eff_{\uu}\partial_x\vartheta_{\uu}=0$, making the convective transport of initial fluctuations apparent.
The initial thermal filling function fluctuates as $\vartheta_{\underline{u}}(x,0)\simeq\bar{\vartheta}_{\underline{u}}+T^{-1/4}\delta\vartheta_{\underline{u}}(x,0)$ where $\delta\vartheta_{\underline{u}}(x,0)$ is a Gaussian white noise with variance $\sigma^2_{\underline{u}}=\chi_{\underline{u}}/(\bar{\rho}^\mathrm{tot}_{\underline{u}})^2$ and $\chi_{\underline{u}}=\bar{\rho}_{\uu}(1-\bar{\vartheta}_{\uu})$. Here and below, $\bar \bullet$ refer to non-fluctuating averaged quantities, and terms with $\delta \bullet$ refer to $O(1)$ zero-mean fluctuating variables. 

Since generically $\bveff_{\uu}\neq0$ initial fluctuations $\delta \vartheta_{\uu}(x,0)$ are transported in time by the linearized GHD equation $\partial_{\Tt}\delta\vartheta_{\uu}+\sqrt{T}\bveff_{\uu}\partial_{\Tx}\delta\vartheta_{\uu}=0$ as
\be\label{eq:fluc_filling}
    \vartheta_{\uu}(x,t)\simeq  \bar{\vartheta}_{\uu}+T^{-1/4}\delta\vartheta_{\uu}(\Tx-\sqrt{T}\bveff_{\uu}\Tt,0)\, .
\ee
\ch{Above, fluctuations in the effective velocity stemming from $\delta \vartheta_u$ can be neglected, as they are subleading in $T$.}
We emphasize that the \emph{diffusive} scale solution Eq.~\eqref{eq:fluc_filling} is fully determined by initial fluctuations and there are no dynamically generated fluctuations as in MFT.

\para{Spin fluctuations.}
 We now connect fluctuations of magnetization with those of giant magnons.
From the microscopic magnetization $\hat{q}_{0;\ell}$ we introduce a coarse-grained version of it, $q_0(x,t)$, whose fluctuations $\delta q_0(x,t)$ can be interpreted classically.
 
We first relate $\delta q_0(\Tx,\Tt)$ to the associated chemical potential $\delta \mu(\Tx,\Tt)$ through the susceptibility matrix $\delta q_0(\Tx,\Tt)=\mathsf{C}_{00}\delta \mu(\Tx,\Tt)$, where $\mathsf{C}_{00}=\sum_\ell\langle \ch{\hat q_{0;\ell}\hat q_{0;0}}\rangle^{\ch{c}}$. Importantly, fluctuations of chemical potentials associated to other charges do not couple to the magnetization at half-filling. Next, fluctuations $\delta \vartheta_{\underline{u}}(x,t)$ of giant magnons are connected to those of $\delta \mu(\Tx,\Tt)$ as \cite{DeNardis2019}
\be\label{eq:magic}
    \delta \mu(x,t)= -\lim_{\mu\to 0^+} \lim_{n\to \infty}\delta\vartheta_{\uu}(\Tx,\Tt) e^{n\mu}/n.
\ee
Note that $ \lim_{n\to\infty}\delta\vartheta_{\uu}(x,t)$ is independent of $u$ \cite{Piroli2017}. \ch{Note that while this identity has been mainly applied to infinite temperature states in previous literature, its validity is based only on the large-n asymptotic behavior of the occupancies, which is temperature-independent.}
We can thus read off the hydrodynamic equation for $\delta \mu$ --and through the susceptibility matrix for $\delta q_0(x,t)$ -- directly from the GHD equations for $\vartheta_{\uu}$ at large $n$, obtaining
$\partial_\Tt \delta q_0+\sqrt{T}v\partial_\Tx \delta q_0=0$, with the spin velocity
$v= \lim_{\mu \to 0^+}\lim_{n\to \infty}\veff_{\uu}$
\cite{Piroli2017}. Crucially $v(x,t)$ is $u-$independent and has zero average $\bar{v}=0$, hence the leading behavior at large $T$ is determined by its fluctuations, stemming from the initial filling fluctuations as
\be\label{eq:vel_fluctuation}
    v(x,t)\simeq \frac{T^{-1/4}}{2\pi\mathsf{C}_{00}}\sum_{\uu}\mathfrak{m}_{\uu}(E')^{\dr}_{\uu}\delta\vartheta_{\uu}(\Tx-\sqrt{T}\bveff_{\uu}\Tt,0),
\ee
where $E'_{\uu}=\p_uE_{\uu}$ and $\mathfrak{m}_{\underline{u}}$ is the response (slope) of the so-called dressed magnetization $m^\mathrm{dr}_{\underline{u}}$ around half-filling, i.e.~$m^\mathrm{dr}_{\underline{u}} = \mathfrak{m}_{\underline{u}}\mu + \mathcal{O}(\mu^2)$, \ch{which follows from the expansion of the dressing operation \cite{DeNardis2019}}. The definition of the dressing operation and the derivation of Eq.~\eqref{eq:vel_fluctuation} are presented in End Matter.
In summary, around half-filling, magnetization fluctuations $q_0(x,t)\simeq T^{-1/4}\delta q_0(x,t)$ are convectively transported along characteristics (``trajectories'' of giant magnons in Fig.~\ref{fig:cartoon}) $\delta q_0(\Tx,\Tt)=\delta q_0(\Tx-X(\Tt), 0)$
where $\dd X(\Tt)/\dd \Tt=\sqrt{T}v(\Tx,\Tt)$ and $X(0)=0$ .
As a final step, we compute the spin current generating function by integrating over the initial fluctuations. 

\para{The typical integrated current distribution.} 
Our primary interest is the probability distribution $\mathcal{P}(J|t)$ of the time-integrated spin current density, which is encoded in the moment generating function $\langle e^{ \lambda \hat{J}(t)}\rangle=\int\dd J\,e^{\lambda J}\mathcal{P}(J|t)$. 
We are specifically interested in the stationary probability distribution of $\hat{J}(t)$ on the timescale of typical fluctuations, $\hat{J}(t)\sim t^{1/{2z}}$, where the dynamical exponent $z=2$ is set by the variance $\langle \hat{J}^2(t)\rangle^c\sim t^{1/{z}}$. We thus define the asymptotic typical probability distribution as $\mathcal{P}_\mathrm{typ}(j)=\lim_{t\to\infty}t^{1/2z}\mathcal{P}(jt^{1/2z}|t)$.
We express the integrated current as the transferred magnetization across the origin, and then pass to coarse grained variables as $J(T)=T^{1/4}\int_0^\infty\dd x\,[\delta q_0(x,T)-\delta q_0(x,0)]$. After using the convective solution for the evolved fluctuations $\delta q_0(x,T)$, we reach the simple form 
\be\label{eq_curr_fluct}
J(T)=T^{1/4}\int_0^{X(1)}\dd x \, \delta q_0(-x,0)\,,
\ee
which we identify with $\hat{J}(T)$ inside correlation functions evaluated at the diffusive scale.
\ch{This expression solidifies the physical interpretation of Fig.~\ref{fig:cartoon}.}
The spin current follows fluctuations of magnetization, or giant magnons, which emanate from the origin and move only due to scatterings with other excitations. The giant magnons' contribution dominates because they are the slowest: their effective velocities in thermal equilibrium decay exponentially with their bare magnetization $n$ as $v_{\underline{u}}\sim e^{-n\eta}$, where $\eta=\cosh^{-1}\Delta$ \cite{de_nardis_diffusion_2019}.
The mechanism is analogous to that in charged single-file cellular automata~\cite{Yoshimura2025}.

The current generating function $\langle e^{\lambda J(T)}\rangle$, is obtained by integrating solely over the initial fluctuations. We express it as a path integral over $\delta\vartheta_{\uu}(\Tx,0)$ and $\delta q_0(\Tx,0)$ independently 
: $\langle e^{\lambda J(T)}\rangle\simeq \frac{1}{2\pi}\int\mathcal{D}\delta q_0(\cdot,0)\mathcal{D}\delta\vartheta(\cdot,0)e^{\lambda J(T)}e^{\mathcal{L}[\delta q_0(\cdot,0),\delta\vartheta(\cdot,0)]}$  where the Lagrangian $\mathcal{L}=\mathcal{L}[\delta q_0,\delta\vartheta]$ is given by
\begin{equation}\label{eq:generating_function}
   \mathcal{L}=-\int\dd x\left[\sum_{\uu}\frac{(\delta\vartheta_{\underline{u}:n<\Lambda}(x,0))^2}{2\sigma^2_{\underline{u}}} +\frac{(\delta q_0(x,0))^2}{2\mathsf{C}_{00}}\right],
\end{equation}
\ch{and $\Lambda$ is a cut-off that is eventually sent to infinity.}
After straightforward Gaussian calculus similar to that performed in Ref.~\cite{Yoshimura2025}, we find that the generating function takes the form $ \langle e^{\lambda J(T)}\rangle\simeq e^{\xi^4}\left(1+\mathrm{erf}\,\xi^2\right)$, with $\xi^2=\lambda^2\sigma\sqrt{T/8}$ where
\be
\sigma^2=\sum_{\underline{u}}(\mathfrak{m}_{\underline{u}})^2\chi_{\underline{u}}|\bveff_{\underline{u}}|.
\label{eq_variance}
\ee
\begin{figure}[t]
    \centering
\includegraphics[width=\linewidth]{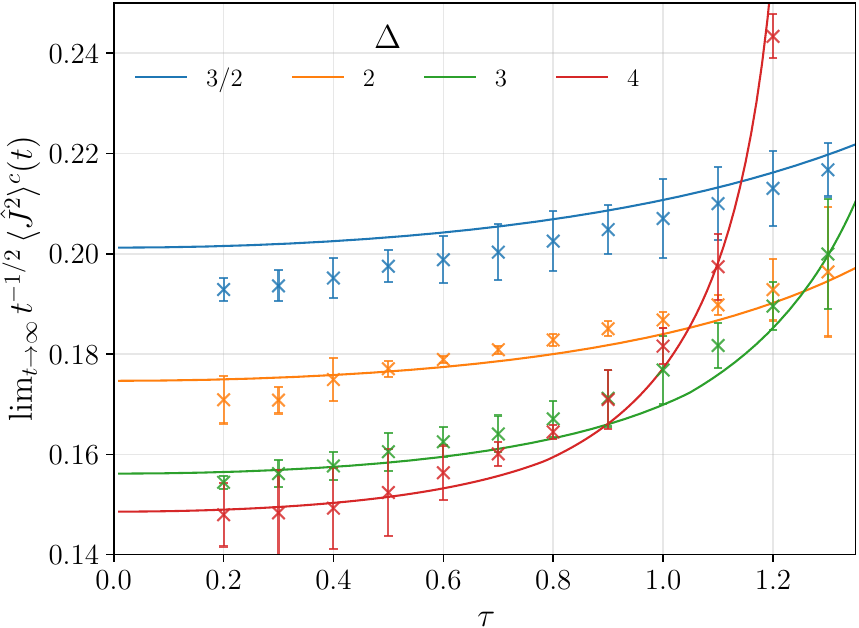}
    \caption{
    \textbf{Variance of the time-integrated spin current}
    in an integrable Trotterization of the XXZ spin chain with time step $\tau$. Numerical data for the spin-current variance (colored crosses) \ch{in the infinite-temperature state} are computed with the quantum generating function approach~\cite{Valli_2025}, and compared with the hydrodynamic prediction $\langle J^2\rangle^c(T)
 \simeq T^{1/2 }\sigma \sqrt{2/\pi}$ where $\sigma^2$ is given by Eq.~\eqref{eq_variance} (solid lines). Error bars show 90\% confidence intervals \ch{stemming from a double extrapolation in time and bond dimension}, see End Matter for details on the numerical simulations.}
    \label{fig:variance}
\end{figure}
Laplace-transforming to the typical probability distribution, we thus obtain the nested Gaussian distribution Eq.~\eqref{eq:typ_dist} reported in the introduction. 
It turns out that the parameter $\sigma^2$ is directly related to the spin diffusion constant $D_\mathrm{s}$ and the spin susceptibility $\mathsf{C}_\mathrm{s}=\mathsf{C}_{00}$ as $\sigma^2=\mathsf{C}_\mathrm{s}^2D_\mathrm{s}$, reproducing the relation found using the ``magic formula'' of Ref.~\cite{DeNardis2019} (see Supplementary Material~\cite{SM}). 
The variance of the current is $\langle J^2(T)\rangle^c
 \simeq \sigma \sqrt{2T/\pi}=\mathsf{C}_\mathrm{s}\sqrt{2TD_\mathrm{s}/\pi}$.
 \ch{While our results are valid at all nonzero temperatures, they require a numerical solution of the thermodynamic Bethe ansatz equations~\cite{Takahashi1999}. For $\beta\to 0$ the solution can be found analytically (see End Matter), leading to $\mathsf{C}_\mathrm{s}=1/4$ and thus $\langle J^2(T)\rangle^c \simeq \sqrt{TD_\mathrm{s}/4\pi}$, in agreement with previous results~\cite{SarangKhemani2024}.}
 
\ch{Although we have exclusively focused on the quantum XXZ chain, the derived formulae involving universal thermodynamic and hydrodynamic objects and thus immediately carry over to other classical and quantum integrable systems. While the nested Gaussian distribution has already been numerically validated in the classical domain \cite{Krajnik2024_1} with great precision, achieving numerical convergence of higher cumulants in the quantum chain has (despite exhibiting anomalous temporal growth) proven much more challenging \cite{Valli_2025,Yadalam2026}. The evidence that Eq.~\eqref{eq:typ_dist} correctly captures the form of spin-current fluctuations is nevertheless quite compelling. It thus remains to quantitatively verify its only free parameter, the variance of Eq.~\eqref{eq:typ_dist}.} To this end, we have performed ab-initio numerical simulations using the integrable Trotterization of the XXZ chain~\cite{Vanicat2017,Ljubotina2019}, and used the quantum generating function approach of Ref.~\cite{Valli_2025}. Details of simulations are reported in End Matter.
 
As shown in Fig.~\ref{fig:variance}, we find satisfactory agreement between the numerically-determined variance and the hydrodynamic prediction. We attribute the larger discrepancy at $\Delta = 3/2$ to stronger finite-time effects due to a superdiffusive transient in the vicinity of the isotropic point $\Delta =1$, in agreement with previous numerical studies~\cite{Prelovsek1,Sirker,Langer,Karrasch2014,BertiniReview2021}.
While higher cumulants show clear indications of non-Gaussian fluctuations, their convergence is very slow and difficult to access through simulations of the quantum model~\cite{Valli_2025,Yadalam2026}. In this respect, we stress once again the importance of classical systems~\cite{Krajnik2024_1} in numerically confirming the form of Eq.~\eqref{eq:typ_dist}.

\para{Discussion}
By employing the BMFT approach of Ref.~\cite{Yoshimura2025}, we have derived the probability distribution of the typical spin current fluctuations in the easy-axis quantum XXZ chain. Our results reveal a common mechanism that gives rise to anomalous fluctuations in both charged single-file systems and the quantum XXZ chain (see also Ref.~\cite{Fujimoto2026} for a microscopic and hydrodynamic computation in the $t0$ model), thereby clarifying the universal origin of this phenomenon.

\ch{The XXZ spin chain is a paradigmatic model for quantum simulators~\cite{Jepsen2020,wei2022,Rosenberg_2024}. Recent advancements in quantum gas microscopes~\cite{wei2022} and superconducting qubits~\cite{Rosenberg_2024} enable in-situ access to the full counting statistics of transferred magnetization.}
\ch{Deterministically prepared Haar-random product states can provide a direct measurement of charge transfer states through repeated measurements. At present, the main bottleneck is arguably the long evolution times required to observe the distinctive form \eqref{eq:typ_dist} of anomalous fluctuations. In this respect, it would be worthwhile exploring quench protocols using certain simple initial states~\cite{Pozsgay2014,Piroli2017Quench}.}

Our work opens up several possible avenues for further investigation. Although here we have focused on typical fluctuations, the BMFT approach can be extended to characterize large fluctuations, which has been achieved in single-file cellular automata \cite{Yoshimura2025}.
It would also be important to apply our predictions to classical integrable spin chains such as the anisotropic Landau-Lifshitz magnet \cite{bastianello2024landaulifschitz}. While the richer quasiparticle spectrum complicates its thermodynamics compared to the infinite-temperature quantum XXZ chain, ab-initio numerical simulations give access to the full current distribution.

\ch{The most pressing issue is applying a similar approach to obtain the structure of anomalous fluctuations in the XXX chain.}
Indeed, spin dynamics at the isotropic point is even more exceptional, exhibiting superdiffusive scaling and an intimate, but not fully understood, connection with the KPZ universality class \cite{Ljubotina2019a,Das2019,Krajnik2020,Weiner2019,Dupont2020,Krajnik2020a,superuniversality,takeuchi2024,muzzi2025}, reviewed in Ref.~\cite{superdiffusion_review}. A microscopic understanding of the fluctuating magnetization would solve this open theoretical challenge, and be of immediate relevance for experiments on quantum simulators~\cite{Scheie2021,wei2022,Rosenberg_2024}.
We foresee that this will require integrating GHD with non-Abelian hydrodynamics~\cite{Lucas2021}, where fluctuations in the polarization of the $SO(3)$-valued magnetization play a role analogous to the magnetization sign considered — or more effectively circumvented — in the present work. 

\para{Acknowledgements.}%
We thank Kazuya Fujimoto, Taiki Ishiyama, Taiga Kurose, Tomohiro Sasamoto, Tomaž Prosen, Marko Žnidarič and Peter Prelovšek for useful discussions and Angelo Valli for providing numerical data used for validation. TY acknowledges financial support from the Royal Society through the University Research Fellowship (\text{URF\textbackslash R1\textbackslash 251593}), and hospitality at the Institute of Science Tokyo.
\v{Z}K is supported by the Simons Foundation as a Junior Fellow of the Simons Society of Fellows (1141511) and thanks the Institute of Science Tokyo for hospitality.
EI is supported by the Research Program P1-0402 and Project N1-0368 funded by the Slovenian Research Agency (ARIS).
\ch{AB is supported by the ERC StG 101220476–NonErgHydro.
Views and opinions expressed are those of the authors only and do not necessarily reflect those of the European Union or the European Research Council Executive Agency. Neither the European Union nor the granting authority can be held responsible for them.}

\para{ Note added.}
Recently, anomalous fluctuations were also explored in different contexts. In Ref.~\cite{Fujimoto2026}, the microscopic as well as hydrodynamic derivations of the exact spin current fluctuations in the $t0$ model, which is the Fermi--Hubbard model with infinitely strong repulsive interactions, were reported. In Ref.~\cite{Urilyon2026-2}, the authors showed that some aspects of anomalous spin current fluctuations in the Heisenberg chain can be captured by a simpler hard-rods-like model by numerically studying its spin current fluctuations.

\bigskip

\section{End Matter}

\subsection*{Thermodynamic Bethe Ansatz}
The Floquet propagator $\mathcal{U}=\mathcal{U}(\tau;\Delta)$ of the Trotterized XXZ circuit with axial anisotropy $\Delta>1$ and timestep $\tau \in \mathbb{R}_{+}$ can be diagonalized using Bethe Ansatz.
The Hamiltonian limit, Eq.~\eqref{eq_Ham}, is retrieved by taking $\tau\to 0$.
Here we only gather the main formulae and refer the reader to Refs.~\cite{Vernier_2023,Zadnik_2024} for details.

The excitation spectrum consists of magnons and their bound states. The dispersion relation of unbound magnons is encoded in the Floquet quasimomentum $p(u)$ and quasienergy $E(u)$, reading
\begin{align}
    e^{-2\ii p(u)} = \frac{\sin{(u_{+}+\ii \eta/2)}}{\sin{(u_{+}-\ii \eta/2)}}
    \frac{\sin{(u_{-}+\ii \eta/2)}}{\sin{(u_{-}-\ii \eta/2)}},\\
     e^{2\ii E(u)} = \frac{\sin{(u_{+}-\ii \eta/2)}}{\sin{(u_{+}+\ii \eta/2)}}
    \frac{\sin{(u_{-}+\ii \eta/2)}}{\sin{(u_{-}-\ii \eta/2)}}.
\end{align}
Circuit anisotropy $\eta$ and spectral shift $\delta=\delta(\tau)$, entering through $u_{\pm} \equiv u\pm \delta/2$, are determined via parameter transmutation relations
\begin{align}
    \cosh \eta = \frac{\sin{(\Delta \tau/2)}}{\sin{(\tau/2)}},\quad
    \sin{(\delta)} = \tan{(\tau/2)} \sinh \eta.
\end{align}
In the thermodynamic limit, $n$-magnon bound states with $n\in \mathbb{N}$ become well-defined asymptotic excitations. Grouping the quantum labels, $\underline{u}\equiv (n,u)$, with $u\in [-\pi/2,\pi/2)$, their quasimomenta and quasienergies, denoted by $p_{\underline{u}}$ and $E_{\underline{u}}$, respectively, are obtained by the fusion of elementary magnons. In terms of $K_{n}(u)\equiv\tfrac{1}{2\pi}\tfrac{2\sinh{(n\eta)}}{\cosh{(n\eta)}-\cos{(2u)}}$, they take the form $p_{\underline{u}}=\tfrac{1}{2}(K_{n}(u_{+})+K_{n}(u_{-}))$
and $E_{\underline{u}}=\tfrac{1}{2}(K_{n}(u_{+})-K_{n}(u_{-}))$.

Equilibrium macrostates are fully specified by the Fermi occupation functions $\vartheta_{\underline{u}}\equiv \rho_{\underline{u}}/\rho^{\rm tot}_{\underline{u}}$, where the quasiparticle rapidity densities $\rho_{\underline{u}}$ and the total state densities $\rho^{\rm tot}_{\underline{u}}(u)$ are related via
\begin{equation}\label{eq:Bethe_Yang}
    \rho^{\rm tot}_{\underline{u}}=\frac{1}{2\pi}\partial_{u} p_{\underline{u}}+
    \mathcal{T}_{\underline{u}}^{~\underline{u'}}\rho_{\underline{u'}},    
\end{equation}
with the summation convention over the repeated discrete index and convolution over the domain of the common rapidity variable in mind.
The scattering kernel $\mathcal{T}_{\underline{u}}^{~\underline{u'}}\equiv -\mathcal{K}_{n,n'}(u-u')$ decomposes as
\begin{equation}
    \mathcal{K}_{n,n'}(u)=\frac{1}{2\pi}\sum_{m=|n-n'|/2}^{(n+n')/2+1}\!(\partial_{u}p_{2m}(u)+\partial_{u}p_{2m+2}(u)).
\end{equation}
Equation \eqref{eq:Bethe_Yang} is physically understood as the dressing of bare momenta $p_{\underline{u}}$ due to elastic scattering with other excitations. Introducing the dressing operation
\begin{equation}
    g_{\underline{u}}^{\rm dr} = [R^{-\mathrm{T}}]_{\underline{u}}^{~\underline{u'}}g_{\underline{u'}},\qquad R \equiv 1-\vartheta\mathcal{T},
\end{equation}
for any dummy functions $g_{\underline{u}}$, Eq.~\eqref{eq:Bethe_Yang} reads simply $2\pi\rho^{\rm tot}_{\underline{u}}=(\partial_{u}p_{\underline{u}})^{\rm dr}=[R^{-\mathrm{T}}]_{\underline{u}}^{~\underline{u'}}\partial_{u}p_{\underline{u}}$.

We consider grand canonical equilibrium states $\hat{\varrho}\simeq e^{\mu \hat{S}^{z}}/Z$ \ch{where dressed quantities can be computed analytically}. The corresponding macrostates, which do not depend on the quasiparticle dispersion laws, read explicitly $\vartheta_{n}(u)=1/\mathcal{X}^{2}_{n}(\mu)$ with $\mathcal{X}_{n}(\mu) \equiv \frac{\sinh{((n+1)\mu/2)}}{\sinh{(\mu/2)}}$. On the other hand, the effective velocities of bound states, $v^{\rm eff}_{\underline{u}}\equiv \partial_{u} E_{\underline{u}}/\partial_{u} p_{\underline{u}}$, do depend on the timestep $\tau$,
\begin{equation}
    v^{\rm eff}_{\underline{u}}(\tau) = \frac{\rho^{(0)\rm tot}_{n}(u_{+})-\rho^{(0)\rm tot}_{n}(u_{-})}{\rho^{(0)\rm tot}_{n}(u_{+})+\rho^{(0)\rm tot}_{n}(u_{-})},
\end{equation}
where $\rho^{(0)\rm tot}_{n}(u)$ denote the total state densities of the \emph{homogeneous} Hamiltonian spin chain, Eq.~\eqref{eq_Ham}, reading
\begin{equation}
    \rho^{(0)\rm tot}_{n}(u) = \frac{\mathcal{X}_{n}(\mu)}{\mathcal{X}_{1}(\mu)}
    \left(\frac{K_{n}(u)}{\mathcal{X}_{n-1}(\mu)}-\frac{K_{n+2}(u)}{\mathcal{X}_{n-1}(\mu)}\right).
\end{equation}
Away from half-filling, i.e.~for $\mu\neq 0$, the second cumulant $c_{2}(T)$ of magnetization transfer (i.e.~the Drude self-weight) exhibits linear asymptotic growth with time $T$, $c_{2}(T)\simeq s_{2}\sqrt{T}$, with
\begin{equation}
    s_{2} = \sum_{\underline{u}}\rho^{\rm tot}_{\underline{u}}\vartheta_{\underline{u}}(1-\vartheta_{\underline{u}})|\veff_{\underline{u}}|(m^{\rm dr}_{\underline{u}}(\mu))^{2},
\end{equation}
where $m^{\rm dr}_{n}(\mu) \equiv \partial_{\mu}\log{(\mathcal{X}^{2}_{n}(\mu)-1)}$ represents the dressed magnetization of the quasiparticles. Since $\lim_{\mu \to 0}m^{\rm dr}_{n}=0$, $s_{2}$ vanishes at half-filling with the curvature
\begin{equation}
    \lim_{\mu\to 0}\partial^{2}_{\mu}s_{2}(\mu) = \sum_{\underline{u}}\rho^{\rm tot}_{\underline{u}}\vartheta_{\underline{u}}(1-\vartheta_{\underline{u}})|\veff_{\underline{u}}|\mathfrak{m}_{\underline{u}}^{2},
\end{equation}
where
\begin{equation}\label{eqn:frakm}
    \mathfrak{m}_{\underline{u}}=\lim_{\mu \to 0}\partial_{\mu}m^{\rm dr}_{\underline{u}}(\mu)=\frac{1}{6}(n+1)^{2}.  
\end{equation}

\subsection{Numerical simulations}
To test our predictions, we have performed extensive numerical simulations of the integrable Trotterization of the XXZ spin chain \cite{Vanicat2017,Ljubotina2019}, with the Floquet propagator
\begin{equation}
    \mathcal{U}(\tau) = \prod_{\ell = 1}^{L/2} U_{2\ell-1, 2\ell}
    \prod_{\ell = 1}^{L/2} U_{2\ell, 2\ell+1},
\end{equation}
with $U_{\ell, \ell+1} = \exp{[-\ii \tfrac{\tau}{4}(\hat{\sigma}^x_\ell\hat{\sigma}^x_{\ell+1}+\hat{\sigma}^y_\ell\hat{\sigma}^y_{\ell+1}+\Delta \hat{\sigma}^z_\ell\hat{\sigma}^z_{\ell+1})]}$ and time step $\tau \in \mathbb{R}_+$.
We used the quantum generating function approach introduced in Ref.~\cite{Valli_2025} which expresses the generating function as a product of operators
\begin{equation}
\langle e^{\lambda \hat{J}(t)} \rangle = 2^{-L} {\rm Tr}\, \left[ \hat \Lambda(\lambda, t) \hat \Lambda(-\lambda, 0)\right], \label{eq:QGF}
\end{equation}
where $\hat \Lambda (\lambda,t=n\tau) = \mathcal{U}^n(\tau) \hat \Lambda(\lambda, 0)\, [\mathcal{U}^n(\tau)]^\dagger$ is a time-evolved operator and $\hat \Lambda(\lambda, 0) = \bigotimes_{\ell = 1}^{L} e^{\lambda_\ell \hat{\sigma}^z_\ell/2}$ with $\lambda_\ell = \lambda \in \mathbb{R}$ for $1 \leq \ell \leq L/2$ and $\lambda_\ell = 0$ otherwise.
The tensor network simulations were performed using the Julia implementation of the ITensor library \cite{itensor,itensor-r0.3} with maximal bond dimension $\chi = 2^8$ on a chain of length $L = 2^8$ for times up to $t_{\rm max}=2^7$. The estimation and extrapolation of the variance of the integrated spin current from the generating function \eqref{eq:QGF} is reported in the Supplementary Material \cite{SM}.

\appendix

\subsection{Fluctuating spin velocity}
We first note that the fluctuating spin velocity $v(x,t)=j_0(x,t)/q_0(x,t)$ is also given by $v(x,t)=\tilde{\jmath}_0(x,t)/\tilde{q}_0(x,t)$ where $\tilde{q}_0=\frac{1}{2}-\sum_{\underline{u}}n\rho_{\underline{u}}$ and $\tilde{\jmath}_0=\sum_{\underline{u}}n\rho_{\underline{u}}\bveff_{\underline{u}}[\rho]$, which follows from the fact that both $q_0$ and $j_0$ include the sign of magnetization, which cancel out in $v$. We first focus on fluctuations of the current $\tilde{\jmath}_0$. Noting that $\tilde{\jmath}_0$ can also be written as $\tilde{\jmath}_0=\sum_{\underline{u}}\frac{n}{2\pi}(E')^{\mathrm{dr}}_{\underline{u}}[\bar{\vartheta}]\vartheta_{\underline{u}}$ and we know that $\vartheta_{\underline{u}}(x,t)$ fluctuates as Eq.~\eqref{eq:fluc_filling}, we only need to understand the way $(E')^{\mathrm{dr}}_{\underline{u}}(x,t)=(E')^{\mathrm{dr}}_{\underline{u}}[\vartheta_\cdot(x,t)]$ fluctuates. Given that
\begin{align}
    &(E')^{\mathrm{dr}}_{\underline{u}}(x,t)\simeq E_{\underline{u}}'+\mathcal{T}_{\underline{u}}^{~\underline{u'}}\bar{\vartheta}_{\underline{u'}}(E')^{\mathrm{dr}}_{\underline{u'}}(x,t)\n
&\quad+T^{-\frac{1}{4}}\mathcal{T}_{\underline{u}}^{~\underline{u'}}\delta\vartheta_{\underline{u'}}(x-\sqrt{\tau}\bveff_{\underline{u'}}t,0)(E')^{\mathrm{dr}}_{\underline{u'}}(x,t),
\end{align}
defining $(E')^{\mathrm{dr}}_{\underline{u}}(x,t)\simeq  ( E')^{\mathrm{dr}}_{\underline{u}}+T^{-1/4}\delta(E')^{\mathrm{dr}}_{\underline{u}}(x,t)$, we have
\begin{align}
\delta(E')^{\mathrm{dr}}_{\underline{u}}&=\mathcal{T}_{\underline{u}}^{~\underline{u'}}(\bar{\vartheta}_{\underline{u'}} \delta(E')^{\mathrm{dr}}_{\underline{u'}}+\delta\vartheta_{\underline{u'}}(x-\sqrt{T}\bveff_{\underline{u'}}t,0)(E')^{\mathrm{dr}}_{\underline{u'}})\n
    &=(\delta\mathcal{E}(x,t))_{\underline{u}}^\mathrm{dr},
\end{align}
where $\delta\mathcal{E}_{\underline{u}}(x,t)=\mathcal{T}_{\underline{u}}^{~\underline{u'}}\delta\vartheta_{\underline{u'}}(x-\sqrt{T}\bveff_{\underline{u'}}t,0)(E')^{\mathrm{dr}}_{\underline{u'}}$. Using these, a simple calculation with the aid of the identity $1+\vartheta \mathcal{T}^\dr=R^{-1}$ yields the fluctuating current $\tilde{\jmath}_0(x,t)$
\begin{equation}
     \tilde{\jmath}_0(x,t)\simeq\frac{T^{-1/4}}{2\pi}\sum_{\underline{u}}n^\mathrm{dr}_{\underline{u}}(E')^{\mathrm{dr}}_{\underline{u}}\delta\vartheta_{\underline{u}}(x-\sqrt{T}\bveff_{\underline{u}}t,0).
\end{equation}
Now we make an important observation: the dressed magnetization, which exactly vanishes at half-filling by symmetry, behaves for small $\mu$ as $m^\mathrm{dr}_{\underline{u}}= \mathfrak{m}_{\underline{u}}\mu + O(\mu^{2})$. For instance, at infinite temperature $\mathfrak{m}_{\underline{u}}$ is given by Eq.~\eqref{eqn:frakm}. Assuming that this replacement is valid in the sum (i.e. the sum over $n$ is convergent upon the replacement), the current $\tilde{\jmath}_0(x,t)$ near $\mu=0$ thus behaves as
\begin{equation}
\tilde{\jmath}_0(x,t)\simeq\frac{T^{-1/4}\mu}{2\pi}\sum_{\underline{u}}\mathfrak{m}_{\underline{u}}(E')^{\mathrm{dr}}_{\underline{u}}\delta\vartheta_{\underline{u}}(x-\sqrt{T}\bveff_{\underline{u}}t,0).
\end{equation}
We generically expect that the linear decrease of $\tilde{\jmath}_0(x,t)$ in $\mu$ near half-filling is precisely canceled out when considering the fluctuating velocity $v=\tilde{\jmath}_0/\tilde{q}_0$, where $\tilde{q}_0$ fluctuates as $\tilde{q}_0(x,t)\simeq\tilde{\mathsf{q}}_0+T^{-1/4}\delta\tilde{q}_0(x,t)$ with $\tilde{\mathsf{q}}_0=\frac{1}{2}-\sum_{\underline{u}}n\rho_{\underline{u}}$. This is because near half-filling $\tilde{\mathsf{q}}_0$ behaves as $\tilde{\mathsf{q}}_0\sim \mathsf{C}_\mathrm{s} \mu$, where $\mathsf{C}_\mathrm{s}$ is the spin susceptibility, by definition. Indeed, again at infinite temperature, we can demonstrate that the density $\tilde{\mathsf{q}}_0$ is given by $\tilde{\mathsf{q}}_0=\tfrac{1}{2}\tanh{(h/2)}$, which implies $\mathsf{C}_\mathrm{s}=1/4$.

Combining these, at half-filling, we conclude that the velocity $v(x,t)$ fluctuates as Eq.~\eqref{eq:vel_fluctuation}.

\subsection{Moment generating function}
To perform the path integral Eq.~\eqref{eq:generating_function}, we first integrate over the initial charge \ch{fluctuation} $\delta q_0(x,0)$. \ch{The relevant part of the path integral is evaluated as (using Eq.~\eqref{eq_curr_fluct})
\begin{align}
    &\frac{1}{2\pi}\int\mathcal{D}\delta q_0(\cdot,0)\exp\left[\lambda J(T) - \int\dd \Tx \frac{(\delta q_0(\Tx,0))^2}{2\mathsf{C}_{00}} \right] \n
    &=e^{\frac{1}{2}\mathsf{C}_{00}\lambda^2T^{1/2}|X(1)|}.
\end{align}}
Plugging this back to Eq.~\eqref{eq:generating_function}, we thus have
\begin{align}\label{eq:theta_integral}
    \langle e^{\lambda J(T)}\rangle&=\frac{1}{\sqrt{2\pi}}\int\mathcal{D}\delta\vartheta(\cdot,0) e^{-\frac{1}{2}\sum_{\underline{u}:n<\Lambda}\int\dd x\,(\delta\vartheta_{\underline{u}}(x,0))^2}\n
    &\quad\times e^{|\sum_{\underline{u}}\int\dd x\,a_{\underline{u}}(x)\delta\vartheta_{\underline{u}}(x,0))|},
\end{align}
where
\begin{equation}
    a_{\underline{u}}(x)=\frac{T^{1/4}\lambda^2}{2}\sqrt{\chi_{\underline{u}}}\mathfrak{m}_{\underline{u}}\mathrm{sgn}(\bveff_{\underline{u}})\mathbf{1}_{A}(x)
\end{equation}
with $A \equiv (-\sqrt{T}|\bveff_{\underline{u}}|,0)$. \ch{Importantly, the sum $\sum_{\underline{u}:n<\Lambda}$ can be now extended to the unrestricted one as the contribution from $n=\infty$ is exponentially suppressed due to $\bveff_{\underline{u}}\sim e^{-n\eta}$.}
The path-integral in Eq.~\eqref{eq:theta_integral} then reduces to a one-dimensional integral \cite{Yoshimura2025}, which reads
\begin{equation}
    \langle e^{\lambda J(T)}\rangle\simeq \frac{1}{\sqrt{2\pi}}\int dX\,e^{-\frac{1}{2}X^2}e^{\sqrt{2}\xi^2|X|}=e^{\xi^4}(1+\mathrm{erf}\,\xi^2)
\end{equation}
where $\xi^2=\lambda^2\sigma\sqrt{T/8}$ and $\sigma^2=\sum_{\underline{u}}(\mathfrak{m}_{\underline{u}})^2\chi_{\underline{u}}|\bveff_{\underline{u}}|$.

\bibliography{bib.bib}
\clearpage
\onecolumngrid
\newpage
\setcounter{equation}{0}  
\setcounter{figure}{0}
\setcounter{page}{1}
\setcounter{section}{0}    
\renewcommand\thesection{\arabic{section}}    
\renewcommand\thesubsection{\arabic{subsection}}    
\renewcommand{\thetable}{S\arabic{table}}
\renewcommand{\theequation}{S\arabic{equation}}
\renewcommand{\thefigure}{S\arabic{figure}}
\setcounter{secnumdepth}{2}  

\begin{center}
{\Large Supplementary Material\\
\vspace{5pt}
Anomalous hydrodynamic fluctuations in the quantum XXZ spin chain
} 
\end{center}
\bigskip
\bigskip
\section{Variance estimation and extrapolation}
\begin{figure}
    \centering
\includegraphics[width=\linewidth]{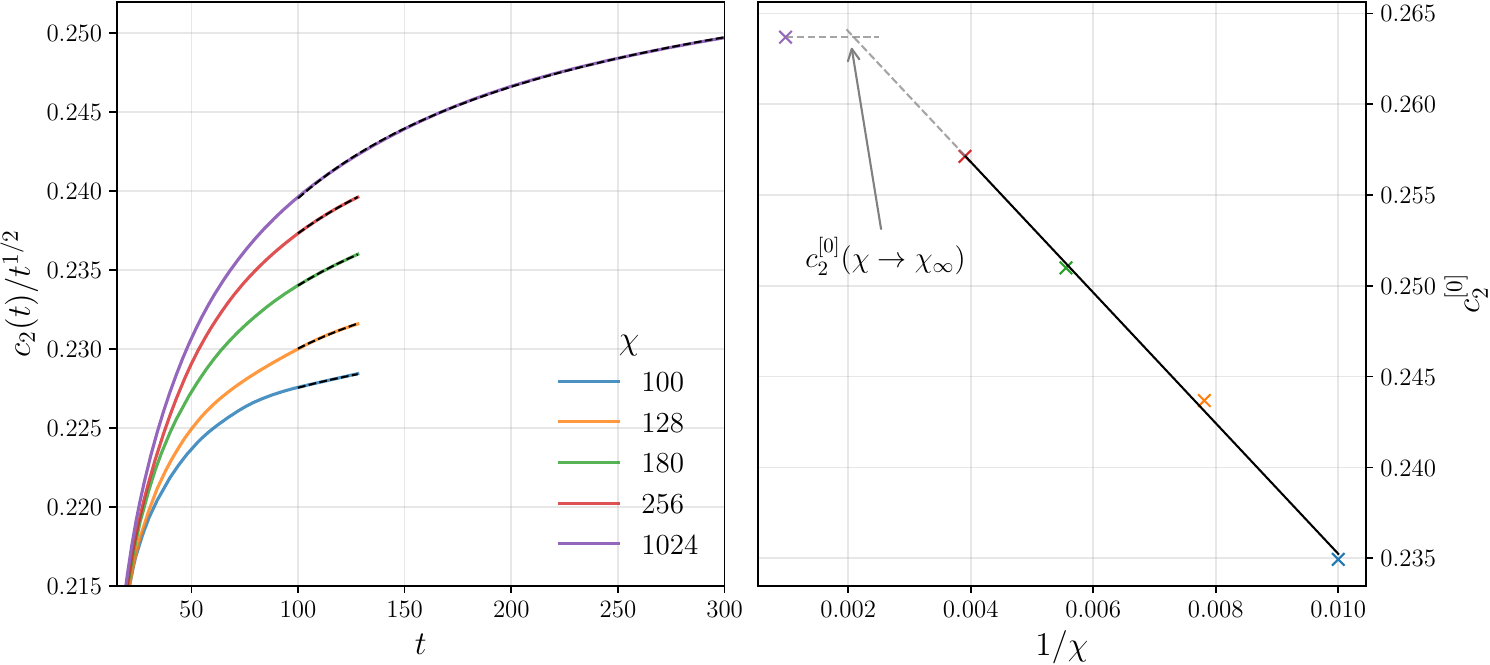}
    \caption{(left panel) Time evolution of rescaled variance of the integrated spin current at different bond dimensions $\chi$ (coloured curves) at $\Delta = 2$, $\tau = 1$. Dashed black lines show the best two-parameter fits \eqref{c_time_fit}. Purple curve shows data from Ref.~\cite{Valli_2025} for longer times and $\chi=1024$, supporting the expansion \eqref{c_time_fit}. (right panel) Time-extrapolated values $c_2^{[0]}$ (coloured crosses) show an approximately linear dependence on $\chi^{-1}$ for intermediate bond dimensions and saturation at the largest bond dimension. Black line show best linear fit \eqref{LS_def} for intermediate bond dimensions. The point of saturation $\chi_\infty$ is estimated by extrapolating to the saturated value.}
    \label{fig:extrapolation}
\end{figure}

In equilibrium, the generating function defined in Eq.~(22) of End Matter is an even function of $\lambda$. The lowest order of its $\lambda$-expansion is quadratic is related to the variance $c_2(t) = \langle\hat{J}^2\rangle^c(t)$
\begin{equation}
    \langle e^{\lambda \hat{J}(t)} \rangle =  1 + c_2(t)\frac{\lambda^2}{2} + \mathcal{O}(\lambda^4).
\end{equation}
We estimate it by evaluating the generating function at $\lambda \ll 1$
\begin{equation}
c_2(t, \chi) \approx 2\lambda^{-2}(\langle e^{\lambda \hat{J}(t)}\rangle - 1),
\end{equation}
where we have additionally emphasized its dependence on the maximal bond dimension $\chi$ of the tensor network.
For all reported simulations we used $\lambda = 10^{-3}$ and verified that this gives converged (in $\lambda$) values of the variance at simulated timescales. To infer its asymptotic value we perform a double extrapolation, first in time, and then in the bond dimension. Since the variance grows asymptotically as $t^{1/2}$, we expand it as a series in falling powers of $t^{1/2}$, fit the finite-time result
\begin{equation}
    c_2(t, \chi) t^{-1/2} = c_2^{[0]}(\chi) +  c_2^{[1]}(\chi)t^{-1/2} + \mathcal{O}(t^{-1})\, \label{c_time_fit}
\end{equation}
and extract $c_2^{[0]}(\chi)$. While there is a priori no theoretical explanation for the $t^{-1/2}$ scaling of the subleading term in \eqref{c_time_fit} this is found to describe the data well, see purple curve in left panel of Fig.~\ref{fig:extrapolation}. We note that the timescale of convergence is not uniform at different anisotropies. In the vicinity of the isotropic point $\Delta = 1$, we expect a crossover from a super-diffusive scaling $c_2 ~ t^{3/2}$ at short times to the diffusive scaling \eqref{c_time_fit} across a timescale $\tau_{\rm crossover} \propto (\Delta - 1)^{-\gamma}$, where $\gamma > 0$. This crossover behavior introduces an additional systematic source of bias into our estimates for smaller $\Delta$.

To account for the dependence of the variance on $\chi$ we further perform a linear least-square fit of $c_2^{[0]}$ against $\chi^{-1}$ for $\chi \in \mathcal{X} = \{100, 128, 180, 256\}$ in which range the relationship is found to be approximately linear and shown in the right panel of Fig.~\ref{fig:extrapolation}
\begin{equation}
    \min_{a, b} \left\{ \sum_{\chi \in \mathcal{X}} \left [c_2^{[0]}(\chi) - (a\chi^{-1}+b)\right]^2 \right\}. \label{LS_def}
\end{equation}
Once the optimal values of $(a_{\rm opt}, b_{\rm opt})$ and their confidence intervals are determined, the final value for the variance is extrapolated to a large bond dimension $\chi_{\infty}$ as
\begin{equation}
    c_2^{[0]}(\chi \to \infty) \approx a_{\rm opt} \chi^{-1}_{\infty} + b_{\rm opt}
\end{equation}
and the corresponding confidence interval computed. While the tensor network faithfully reproduces the full many-body Hilbert space only for $\chi_{\infty} = 2^L$, one finds that expectation values of local quantities converge at much smaller bond dimensions since the relevant states span only a small subset of the full space. Numerical simulations suggest saturation at $\chi_\infty \approx 512$ (see right panel of Fig.~\ref{fig:extrapolation}) at given system sizes and times which we used in extrapolating the reported values in Fig.~2 of the Main Text.

\section{Spin diffusion constant}
We define the Onsager matrix via the Green-Kubo formula
\begin{equation}\label{eq:onsager}
    L_{ij}=\int_\mathbb{R}\dd t\left(\int_\mathbb{R}\dd x\langle \hat{\jmath}_i(x,t)\hat{\jmath}_j(0,0)\rangle^c-\mathsf{D}_{ij}\right),
\end{equation}
where $\mathsf{D_{ij}}=\lim_{t\to\infty}\int_\mathbb{R}\dd x\langle \hat{\jmath}_i(x,t)\hat{\jmath}_j(0,0\rangle^c$ is the Drude weight. The Einstein relation then relates this to the diffusion matrix through $L=D\mathsf{C}$. To compute the spin diffusion constant $D_\mathrm{s}=D_{00}$ at half-filling where the spin Drude weight $\mathsf{D}_{00}$ vanishes, we first note that $\mathsf{C}_{i0}=\mathsf{C}_\mathrm{s}\delta_{0i}$ holds there, implying ${L}_{00}=D_\mathrm{s}\mathsf{C}_\mathrm{s}$. We therefore need to evaluate ${L}_{00}$ only. While it is straightforward to directly compute Eq.~\eqref{eq:onsager}, we shall deal with another equivalent representation of it
\begin{equation}
  L_{00}= \lim_{t\to\infty}\frac{1}{t}\int_\mathbb{R}\dd x\,x^2\langle \hat{q}_0(x,t)\hat{q}_0(0,0)\rangle^c,
\end{equation}
as we wish to demonstrate how the spin-spin correlator $\langle \hat{q}_0(x,t)\hat{q}_0(0,0)\rangle^c$ behaves at the diffusive scale first.
To this end, we focus on evaluating $\langle q_0(x,t) q_0(0,0)\rangle^c$, in terms of which the spin Onsager coefficient reads
\begin{equation}\label{eq:ons}
   L_{00}=  \lim_{T\to\infty}\sqrt{T}\lim_{t\to\infty}\frac{1}{t}\int_\mathbb{R}\dd x\,x^2\langle q_0(x,t) q_0(0,0)\rangle^c.
\end{equation}
Let us start with noticing that one path-integral involved in evaluating $\langle q_0(x,t) q_0(0,0)\rangle^c$ can be performed immediately: since we have $\langle q_0(x,t) q_0(0,0)\rangle=T^{-1/2}\langle \delta q_0(x-X(t),0) \delta q_0(0,0)\rangle$, the path-integral over $\delta q_0(x,0)$ can be taken first using $\langle \delta q_0(x,0) \delta q_0(y,0)\rangle=\mathsf{C}_\mathrm{s}\delta(x-y)$, yielding
\begin{equation}
    \label{eq:qq_asymp1}
    \langle q_0(x,t) q_0(0,0)\rangle^c \simeq \frac{T^{-\frac{1}{2}}\mathsf{C}_\mathrm{s}}{(2\pi)^\frac{3}{2}}\int\mathcal{D}\delta\vartheta(\cdot,0)\int\dd k\,e^{\ii k(x-X(t))}\exp\left[-\sum_{\underline{u}}\int\dd x\frac{(\delta\vartheta_{\underline{u}}(x,0))^2}{2\sigma^2_{\underline{u}}}\right].
\end{equation}
Plugging $X(t)\simeq \frac{T^{-1/4}}{2\pi\mathsf{C}_\mathrm{s}}\sum_{\underline{u}}\mathfrak{m}_{\underline{u}}(p')^\mathrm{dr}_{\underline{u}}\int_{-\sqrt{T}t\bar{v}^\eff_{\underline{u}}}^0\dd x\,\delta\vartheta_{\underline{u}}(x,0)$ into Eq.~\eqref{eq:qq_asymp1}, trivial Gaussian integrals give
\begin{equation}
     \langle q_0(x,t) q_0(0,0)\rangle^c\simeq \frac{T^{-\frac{1}{2}}\mathsf{C}_\mathrm{s}}{\sqrt{2\pi D_\mathrm{s}t}}e^{-x^2/2D_\mathrm{s}t},
\end{equation}
where $D_\mathrm{s}=\mathsf{C}^{-2}_\mathrm{s}\sum_{\underline{u}}\mathfrak{m}^2_{\underline{u}}\chi_{\underline{u}}|\bar{v}^\eff_{\underline{u}}|=\mathsf{C}^{-2}_\mathrm{s}\sigma^2$ is identified with the spin diffusion constant. We can readily verify it by evaluating Eq.~\eqref{eq:ons}, which confirms $L_{00}=D_\mathrm{s}\mathsf{C}_\mathrm{s}$ as anticipated.

\end{document}